\documentclass{article}

\usepackage{amssymb,amsthm}

\newcommand{\X}{\mathsf{X}}     %next
\newcommand{\U}{{\,\uU\,}}      %until
\newcommand{\uU}{\mathsf{U}}    %until without spaces
\newcommand{\uR}{\mathsf{R}}    %release without spaces
\newcommand{\R}{{\,\uR\,}}      %release
\newcommand{\F}{\mathsf{F}}     %in future
\newcommand{\G}{\mathsf{G}}     %globally
\newenvironment{quotenv}[1]{\par\medskip{{\bfseries\noindent#1.}}\it}{\par\medskip}

\begin{document}

\title{A Short Story of a Subtle Error in LTL Formulas Reduction and Divine
  Incorrectness\thanks{Authors have been supported by the Czech Science
    Foundation grant No.~201/09/1389 (Tom{\'a}{\v{s}} Babiak and Mojm\'{\i}r
    K{\v{r}}et{\'{\i}}nsk{\'{y}}), grant No.~201/08/P459 (Vojt\v{e}ch
    {\v{R}}eh\'{a}k), and grant No.~201/08/P375 (Jan Strej{\v{c}}ek).} }

\author{Tom\'{a}\v{s} Babiak \and Mojm\'{\i}r K\v{r}et\'{\i}nsk\'{y} \and
  Vojt\v{e}ch {\v{R}}eh\'{a}k \and Jan Strej\v{c}ek \\[5mm]
  Faculty of Informatics, Masaryk University,\\
    Botanick\'a 68a, Brno, Czech Republic.\\
  \texttt{\small \{xbabiak,kretinsky,rehak,strejcek\}@fi.muni.cz}}

\maketitle

  \begin{abstract}
    We identify a subtle error in LTL formulas reduction method used as one
    optimization step in an LTL to B\"uchi automata translation. The error
    led to some incorrect answers of the established model checker
    DiVinE. This paper should help authors of other model checkers to avoid
    this error.
  \end{abstract}

A translation of \emph{Linear Temporal Logic (LTL)} formulas into language
equivalent B\"uchi automata is an important part of all LTL model
checkers. The translation is exponential in the length of the translated
formula. As the size and shape of the produced automaton can greatly affect
running time of other parts of the model checking algorithms, many
improvements of standard translations emerged.
Some of the improvements modify an input LTL formula in order to reduce its
size and number of modal operators. Unfortunately, the modification suggested
in~\cite{EH00} contains an error: it can produce a smaller but non-equivalent formula.

The error is in the definition of \emph{pure eventuality} formulas. 

\begin{quotenv}{Quotation of Definition 2 of \cite{EH00}}
  The class of \emph{pure eventuality} formulas are defined as the smallest
  set of LTL formulas (in negation normal form) satisfying:
  \begin{itemize}
  \item Any formula of the form $\F\varphi$ is a pure eventuality formula.
  \item Given pure eventuality formulas $\psi_1$ and $\psi_2$, and $\gamma$
    an arbitrary formula, each of
    $\psi_1\vee\psi_2$, $\psi_1\wedge\psi_2$, $\psi_1\U\gamma$, $\G\psi_1$,
    $\psi_1\R\psi_2$, and $\X\psi_1$ is also a pure eventuality formula.
  \end{itemize}
\end{quotenv}

The paper~\cite{EH00} claims that all pure eventuality formulas define
left-append closed languages, where a language $L$ is \emph{left-append
  closed} if for all $w\in\Sigma^\omega$ and $v\in\Sigma^*$: if $w\in L$,
then $vw\in L$. One can easily disprove this claim. For example,
$\varphi=(\F b)\U c$ is a pure eventuality formula and $L(\varphi)$ is not
left-append closed as $c^\omega\in L(\varphi)$ and $a.c^\omega\not\in
L(\varphi)$.

Invalidity of the claim causes invalidity of the \emph{Basic Operator
  Reduction Lemma}, which directly describes the reduction steps. We recall
only the part of the lemma related to pure eventuality formulas.

\begin{quotenv}{Quotation of Lemma 3 (Basic Operator Reduction Lemma)
  of~\cite{EH00}, Item 4}
For all LTL formulas $\varphi$ and pure eventuality formulas $\psi$, the
following equivalences hold: $(\varphi\U\psi)\equiv\psi$ and
$F\psi\equiv\psi$.
\end{quotenv}

Using the reduction lemma, one can reduce the formula $a\U((\F b)\U c)$ into
$(\F b)\U c$. However, the formulas are not equivalent as $a.c^\omega\models
a\U((\F b)\U c)$ while $a.c^\omega\not\models(\F b)\U c$. Similarly, $\F((\F
b)\U c)$ can be reduced into a non-equivalent formula $(\F b)\U c$. In
general, only the implications $(\varphi\U\psi)\Longleftarrow\psi$ and
$\F\psi\Longleftarrow\psi$ hold. Hence, if an LTL to B\"uchi automata
translation employs this reduction, then there can be a word satisfying an
input formula but not accepted by the resulting automaton. In the context of
model checking, input formulas represent incorrect behaviours. Thus, the
resulting automaton can represent a smaller set of incorrect behaviours than
the input formula specifies. As a result, a model checker with such a
translation can state that a system is correct even if it is not.

We have detected exactly this kind of error in all versions of the model
checker \emph{DiVinE}~\cite{BBC+06} developed during the last five years,
i.e.~DiVinE version~2.2 and DiVinE Cluster version 0.8.2 and all older
versions. The bug has been fixed with our assistance. The fix will appear in
the upcoming versions of DiVinE family tools.

\bigskip Incorrectness of the claim is caused by the part of the
definition saying that, for a pure eventuality formula $\psi_1$ and an
arbitrary formula $\gamma$, $\psi_1\U\gamma$ is also a pure eventuality
formula. To fix it, it is sufficient to replace $\psi_1\U\gamma$ by
$\gamma\U\psi_1$. The proof is straightforward.

A careful researcher can found that on Etessami's web page, there is
a~reference to~\cite{EH00} leading to a PostScript file~\cite{EH00-second},
which is a slightly different version of~\cite{EH00}.\footnote{As the
  differences between~\cite{EH00} and~\cite{EH00-second} are minor
  and~\cite{EH00-second} does not contain any reference to the conference
  version, one tends to think that~\cite{EH00-second} is a preprint
  of~\cite{EH00} rather than a full version.} In~\cite{EH00-second}, the
definition of pure eventuality formulas is repaired in the following way:

\begin{quotenv}{Quotation of Definition 2 of \cite{EH00-second}}
  The class of \emph{pure eventuality} formulas are defined as the smallest
  set of LTL formulas (in negation normal form) satisfying:
  \begin{itemize}
  \item Any formula of the form $\F\varphi$ is a pure eventuality formula.
  \item Given pure eventuality formulas $\psi_1$ and $\psi_2$, each of
    $\psi_1\vee\psi_2$, $\psi_1\wedge\psi_2$, $\psi_1\U\psi_2$, $\G\psi_1$,
    $\psi_1\R\psi_2$, and $\X\psi_1$ is also a pure eventuality formula.
  \end{itemize}
\end{quotenv}

Here, the set of pure eventuality formulas is strictly smaller than the one
defined in \cite{EH00} and, in context of this new definition, the mentioned
claim holds (in fact, \cite{EH00-second} contains a proof). Consecutively,
also Basic Operator Reduction Lemma is correct in this setting.

We note that the set of pure eventuality formulas according to
Definition 2 of \cite{EH00-second} is significantly smaller than the one
obtained by the mentioned replacement of $\psi_1\U\gamma$ by
$\gamma\U\psi_1$. In spite of this, the reduction of LTL formulas presented
in \cite{EH00-second} is not weaker. The reason is that the Basic Operator
Reduction Lemma allows to reduce all the formulas of the form
$\gamma\U\psi_1$ to $\psi_1$. Hence, the final effect of the reduction is
the same in both cases.


\begin{thebibliography}{1}

\bibitem{BBC+06}
J.~Barnat, L.~Brim, I.~{\v{C}}ern\'{a}, P.~Moravec, P.~Ro{\v{c}}kai, and
  P.~{\v{S}}ime{\v{c}}ek.
\newblock {DiVinE} -- a tool for distributed verification (tool paper).
\newblock In {\em Proceedings of {CAV} 2006}, volume 4144 of {\em LNCS}, pages
  278--281. Springer, 2006.

\bibitem{EH00}
K.~Etessami and G.~J. Holzmann.
\newblock Optimizing {B}{\"u}chi automata.
\newblock In {\em Proceedings of {CONCUR} 2000}, volume 1877 of {\em LNCS},
  pages 153--167. Springer, 2000.

\bibitem{EH00-second}
K.~Etessami and G.~J. Holzmann.
\newblock Optimizing {B}{\"u}chi automata.
\newline\texttt{http://homepages.inf.ed.ac.uk/kousha/opting\_buchi\_rucnoc.ps}

\end{thebibliography}
\end{document}